\documentclass[9pt,twocolumn,twoside]{osajnl}

\journal{ol} % Choose journal (ao, aop, josaa, josab, ol)

\setboolean{shortarticle}{true} 
\ifthenelse{\boolean{shortarticle}}{\colorlet{color2}{color2b}}{\colorlet{color2}{color2}} % Automatically switches colors for short articles

\title{Passive optical gyroscope with double homodyne readout}

\author[1,2*]{Denis Martynov}
\author[1]{Nicolas Brown}
\author[1]{Eber Nolasco-Martinez}
\author[1]{Matthew Evans}

\affil[1]{LIGO Laboratory, Massachusetts Institute of Technology, Cambridge,
Massachusetts, 02139, USA}
\affil[2]{School of Physics and Astronomy, and Institute of Gravitational Wave Astronomy, 
University of Birmingham, Edgbaston, Birmingham B15 2TT, United Kingdom}

\affil[*]{Corresponding author: dmartynov@star.sr.bham.ac.uk}

\dates{Compiled \today}

\ociscodes{(060.2800) Gyroscopes; (140.3370) Laser Gyroscopes; (120.5790) Sagnac
effect}

\doi{\url{http://dx.doi.org/10.1364/ao.XX.XXXXXX}}

\usepackage{graphicx}
\usepackage{lineno} %used to add line numbers to each line
\newcommand{\sens}{rad/s$/\sqrt{\rm Hz}$ }
\begin{abstract}
We present a passive, resonant, single-frequency gyroscope design that utilises
polarisation modes of an optical cavity to readout rotation and generate a laser
frequency discriminant.
This design is notable for its simplicity, requiring no
modulation electronics or frequency counters. We extract both the cavity length signal and rotation signal from two co-propagating beams with orthogonal polarisations. This readout
scheme can be applied to an optical cavity whose
polarisation eigen-modes experience different phase shifts such as fibre rings;
whispering gallery mode resonators; and folded free-space cavities. We apply this
technique to the passive free-space gyroscope and achieve a bias stability of 0.03$^\circ$/h and a sensitivity of $5 \times 10^{-8}$\,\sens
above 1\,Hz, with a cavity of area 400\,cm$^2$ and finesse of $10^4$.
Below 1\,Hz the sensitivity of the gyroscope is limited by the backscattering in the optical cavity and beam jitter of the laser beam.
\end{abstract}

\setboolean{displaycopyright}{true}

\begin{document}

\maketitle
\thispagestyle{fancy}

\ifthenelse{\boolean{shortarticle}}{\ifthenelse{\boolean{singlecolumn}}{\abscontentformatted}{\abscontent}}{}

Optical gyroscopes measure rotation of a particular frame using two counter-propagating laser beams. The beams follow the same closed path in the rotating frame and experience different phase shifts according to the Sagnac effect~\cite{Post:67, Chow_1985}. This shift is given by the equation
\begin{equation}\label{eq:dphi}
	\Delta \phi = \frac{8 \pi A \Omega}{c \lambda},
\end{equation}
where $A$ is the area enclosed by the optical beam path, $\Omega$ is the angular frequency of the rotating frame, $c$ is the speed of light, and $\lambda$ is the wavelength of the laser beam. Optical gyroscopes measure the phase or frequency difference between the two counter-propagating beams and infer the frame's rotation speed.

Optical gyroscopes have a wide range of applications (see~\cite{Passarp:17} for a recent review). Active ring laser gyroscopes have a lasing media inside their optical cavities which generates two counter-propagating beams for rotation sensing. $\sim 10$\,cm scale active gyroscopes with bias stability of $<10^{-3}\,^\circ$/hr and angular random walk (ARW) $<10^{-3}\,^\circ/\sqrt{\rm Hr}$, such as the Honeywell GG1320, found application in aeronautics and submarine navigation~\cite{Passarp:17}. Even larger active optical gyroscopes, such as UG-2~\cite{Hurst_2009} and GP-2~\cite{Beverini:16}, study geophysics and an effect of the Moons gravity on the Earth's rotation axis. Passive free-space~\cite{Ezekiel:77} and interferometric fibre optic gyroscopes~\cite{Vali:76, Bergh_1984} use an external laser beam which is split along the two counter-propagating paths. Fibre gyroscopes, such as KVH DSP-1750 and Northrop Grumman LN-200, found applications in platform stabilisation and navigation industries~\cite{Passarp:17}; they generally have similar angular random walk (ARW) to active ring laser gyroscopes but with worse bias stability $<0.1^\circ/$hr. Recently, compact active~\cite{Li_2017} and passive~\cite{Matsko_2018} gyroscopes based on the whispering-gallery-mode resonators were presented. These gyroscopes have a potential to be used in navigation in the future once their bias stability is improved.

Medium size ($\sim 1$\,m) gyroscopes~\cite{Belfi:12, Korth:15} have a potential to significantly improve seismic isolation systems in gravitational-wave detectors, such as LIGO~\cite{LSC_aLIGO_2015} and Virgo~\cite{Acernese_aVIRGO_2015}. In these detectors, test masses are suspended from optical tables which move due to the seismic disturbances. The motion of the LIGO optical tables is actively stabilised in six degrees of freedom~\cite{Matichard2015273} using a set of seismometers or geophones. These inertial sensors have sensitivity of $10^{-10} - 10^{-11}$\,m/$\sqrt{\rm Hz}$ around 1\,Hz. However, at lower frequencies $f \lesssim 1$\,Hz the noise grows as $1/f^2$ and even faster due to mechanical response of suspended masses in these sensors and tilt-to-horizontal coupling~\cite{Venkateswara_2014}. Gyroscopes do not suffer from this problem and can improve inertial isolation of the optical tables in the frequency band from 10\,mHz up to 10\,Hz. Active suppression of ground vibrations can enhance the low-frequency sensitivity of gravitational-wave detectors and make them more sensitive to intermediate-mass black holes~\cite{Yu_5Hz_2018}.

In this letter, we present a passive optical gyroscope with a high-finesse optical cavity and double homodyne readout. This design is free from lock-in behaviour and does not require mechanical dithering~\cite{Wang_2001} to measure small rotation velocities. Such a mechanism can inject noise in the motion of the LIGO optical benches which couples to the gravitational-wave channel via backscattering~\cite{Martynov_THESIS_2015}. 
%However, we note that backscattering in the optical cavity of the gyroscope limits its sensitivity below 1\,Hz.
Since the previous design of the passive gyroscope for the gravitational-wave detectors was limited by electronics noise of the frequency counter~\cite{Korth:15}, we developed a novel readout scheme which employs only homodyne readouts both for stabilising the laser and reading out the frame rotation.

Our scheme avoids radio-frequency optical beats by using two orthogonal polarisation eigenmodes of the folded resonant cavity~\cite{Libson:15}. These eigen-modes acquire different phases upon a non-normal reflection from the cavity mirrors and, therefore, have non-degenerate resonance frequencies. Only one linear polarisation resonates in the optical cavity and contains information about the frequency noise and the frame rotation. The orthogonal polarisation follows the same optical path but does not resonate in the cavity and behaves as a local oscillator in the homodyne readouts. This scheme can be applied to a wide range of sensors such as whispering-gallery-mode gyroscopes, fibre rings and folded free-space cavities.

\begin{figure}
        \includegraphics[width=0.9\columnwidth]{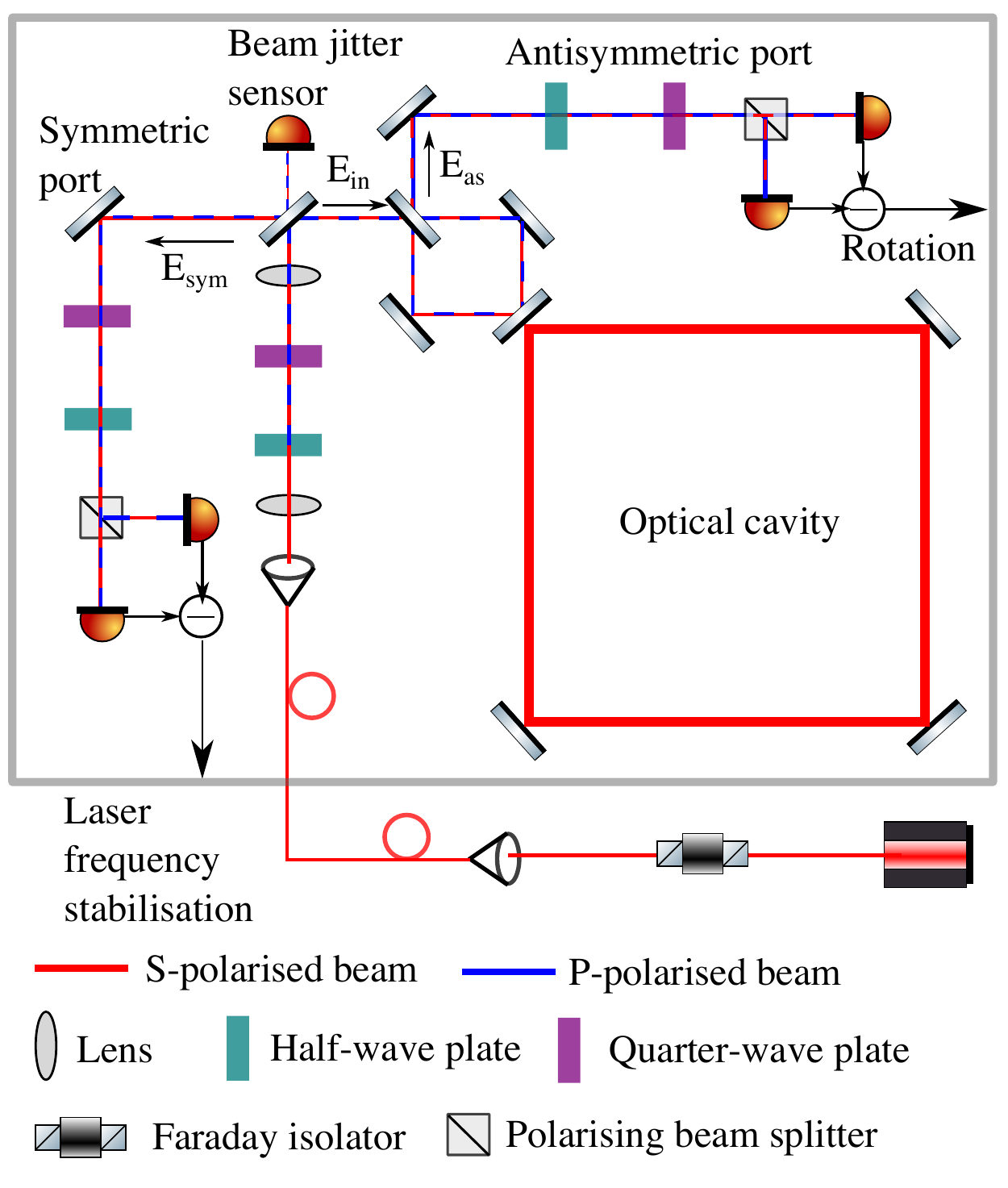}
        \caption{The layout of the optical gyroscope with a double homodyne readout. 
        %The rotation signal is detected at the antisymmetric port.
        The gyroscope is installed in the polystyrene box to passively filter acoustic and thermal fluctuations of the environment.
        %The laser light is coupled inside the box through an optical fibre.
        }
        \label{fig:diagram}
\end{figure}

The optical layout of the gyroscope is shown in Fig.~\ref{fig:diagram}. The input laser beam, with wavelength of $\lambda=1064$\,nm in S-polarisation, passes through a Faraday isolator and couples into an optical fibre. The optical cavity, auxiliary optics and photodetectors are installed in air inside an isolation box. Inside the box, the light passes through mode matching optics, and a set of waveplates which converts vertical polarisation to circular. A fraction of the laser beam is picked off by an input mirror and hits the quadrant photodetector which measures the spatial motion of the beam. The rest of the light is split into two beams which are injected clockwise and counterclockwise into the high-finesse optical cavity of side length $a = 20$cm. Since the cavity is folded and has a planar axis, its eigen-modes have S- and P- polarisations. In our setup, we resonate S-polarisation in the optical cavity and use co-propagating field in the P-polarisation as a local oscillator for the homodyne readouts.
%The reflected light with phase fluctuations common to both the clockwise and counter-clockwise beams returns toward the laser and the symmetric port of the interferometer. Laser light in S-polarisation does not propagate towards the antisymmetric port of the interferometer unless the two counter-propagating beams acquire differential phase shifts due to the frame rotation.
%Reflected cavity light with differential phase shift interferes at the beamsplitter and proceeds toward the antisymmetric port of the interferometer. 
Parameters of the experiment are shown in Table~\ref{table:parameters}.

The beam splitter is balanced (50/50) for S-polarisation and has a power transmission of $T_{\rm bs, p}=0.63$ for P-polarisation. This imbalance leads to the local oscillator fields at the detection ports given by the equations
\begin{equation}\label{eq:p_pol}
\begin{split}
	& E_{\rm sym, p} = 2 \sqrt{T_{\rm bs, p} R_{\rm bs, p}} E_{\rm in,p} \sqrt{T_{\rm in, p}} \\
        & E_{\rm as, p} = (T_{\rm bs, p} - R_{\rm bs, p}) E_{\rm in, p},
\end{split}
\end{equation}
where $E_{\rm in,p}$ , $E_{\rm sym, p}$ and $E_{\rm as, p}$ are P-polarised fields near the beam splitter as shown in Fig.~\ref{fig:diagram}, $T_{\rm in, p}=0.5$ is transmission of the input mirror for the P-polarised beam, and $R_{\rm bs, p} = 1 - T_{\rm bs, p} = 0.37$ is reflectivity of the beam splitter for P-polarisation.

The laser field in the S-polarisation is sensitive to the longitudinal and rotational motion of the optical cavity according to the equations:
\begin{equation}\label{eq:s_pol}
\begin{split}
        & E_{\rm sym, s} = e^{i B\phi_+} \cos(B \frac{\Delta \phi}{2}) E_{\rm in,s} \sqrt{T_{\rm in, s}} \\
        & E_{\rm as, s} = e^{i B\phi_+} i \sin(B \frac{\Delta \phi}{2}) E_{\rm in, s},
\end{split}
\end{equation}
where $B$ is the power build-up factor in the optical cavity, $T_{\rm in, s}=0.3$ is transmission of the input mirror for the S-polarised beam, $E_{\rm in, s}$ , $E_{\rm sym, s}$ and $E_{\rm as, s}$ are S-polarised fields near the beam splitter, and $\phi_+=8\pi a \Delta f/c$ is the common phase shift of the clockwise and counterclockwise beams in the optical cavity due to the difference $\Delta f$ between the cavity eigen-frequency and the laser frequency. The differential phase shift $\Delta \phi$ is given by the Eq.~\ref{eq:dphi}.

 \begin{table}%[H] add [H] placement to break table across pages
 \caption{Parameters of the experiment.}
 \label{table:parameters}
 \begin{tabular}{lc}
 \hline
Cavity area, $A$   & 400\,cm$^2$ \\
Cavity build-up for S-polarisation, B & 6500 \\
Cavity build-up for P-polarisation & 55 \\
Beam splitter transmission, $T_{\rm bs, p}$ & 0.63 \\
Input power in S-polarisation & 4\,mW \\
Input power in P-polarisation & 16\,mW \\
\hline
 \end{tabular}
 \end{table}

The output waveplates and polarising beam splitters at the symmetric and antisymmetric ports rotate the resonant and non-resonant light, mixing the two, so that the differential signals are $K_{\rm sym} = 2 \Re ( E_{\rm sym,p} E_{\rm sym, s}^*)$ and $K_{\rm as} = 2 \Re ( E_{\rm as,p} E_{\rm as,s}^*)$. Therefore, if the output fields in S- and P- polarisations are orthogonal in the complex plane then homodyne signals are zero. Eqs.~\ref{eq:p_pol} and~\ref{eq:s_pol} imply that $K_{\rm sym}=0$ if $\phi_+ = 0$ since the input light is circular polarised.
%If $\phi_+ \neq 0$ then we get signal in the symmetric port.
In the linear regime ($B \phi_+ \ll 1$), the homodyne signal is given by the equation
\begin{equation}\label{eq:cal_sym}
\begin{split}
        & \frac{K_{\rm sym}}{\Delta f}= 32\pi \sqrt{T_{\rm bs, p} R_{\rm bs, p}} \sqrt{T_{\rm in, p}T_{\rm in, s}} \sqrt{P_{\rm in, p} P_{\rm in, s}} \cdot B \frac{a}{c} \\
	& = 6.6 \cdot 10^{-7} 
		\sqrt{ \frac{P_{\rm in, p}}{16\,{\rm mW}}
		 \frac{P_{\rm in, s}}{4\,{\rm mW}}}
		 \left( \frac{a}{20\,{\rm cm}} \right)
        	\left( \frac{B}{6500} \right) \frac{\rm W}{\rm Hz}.
\end{split}
\end{equation}

The field $E_{\rm as, s} \neq 0$ only when the gyroscope rotates with a non-zero angular velocity. For small phase shifts ($B \phi_- \ll 1$), the calibration of the gyroscope is given by the equation
\begin{equation}\label{eq:cal_as}
\begin{split}
	& \frac{K_{\rm as}}{\Omega} = 8 \pi (T_{\rm bs, p} - R_{\rm bs, p}) \sqrt{P_{\rm in, p} P_{\rm in, s}} B \frac{A}{c \lambda} \\
	& = 4.6\cdot 10^{-2}
		\sqrt{ \frac{P_{\rm in, p}}{16\,{\rm mW}}
		 \frac{P_{\rm in, s}}{4\,{\rm mW}}}
		 \left( \frac{a}{20\,{\rm cm}} \right)^2
		 \left( \frac{B}{6500} \right)
		 \frac{\rm W}{\rm rad/s}. \\
\end{split}
\end{equation}

\begin{figure}[!t]
        \includegraphics[width=0.9\columnwidth]{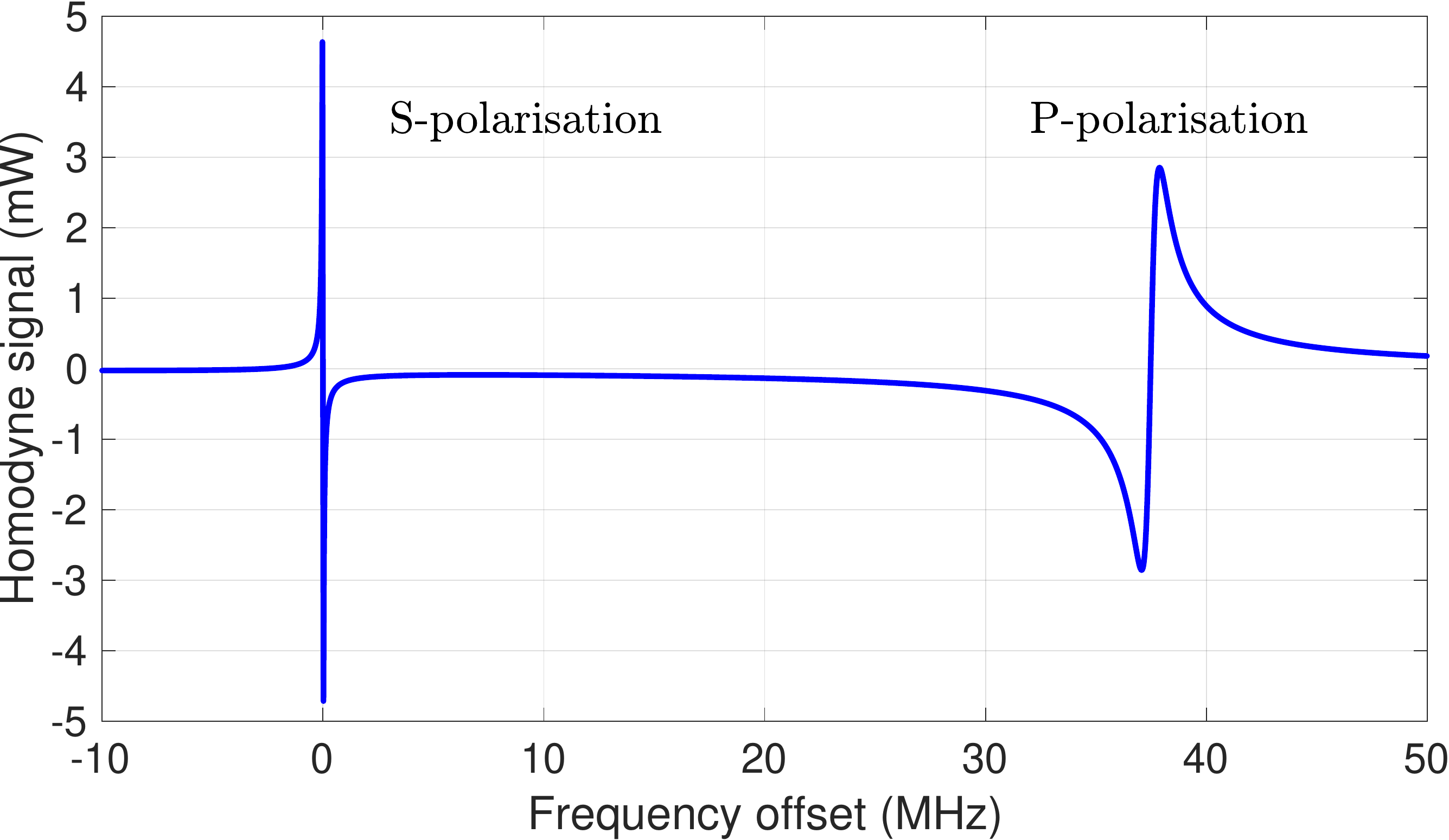}
        \caption{Response of the gyroscope to the different between the cavity length and the laser frequency.}
        \label{fig:homodyne}
\end{figure}

\begin{figure}[!t]
        \includegraphics[width=0.9\columnwidth]{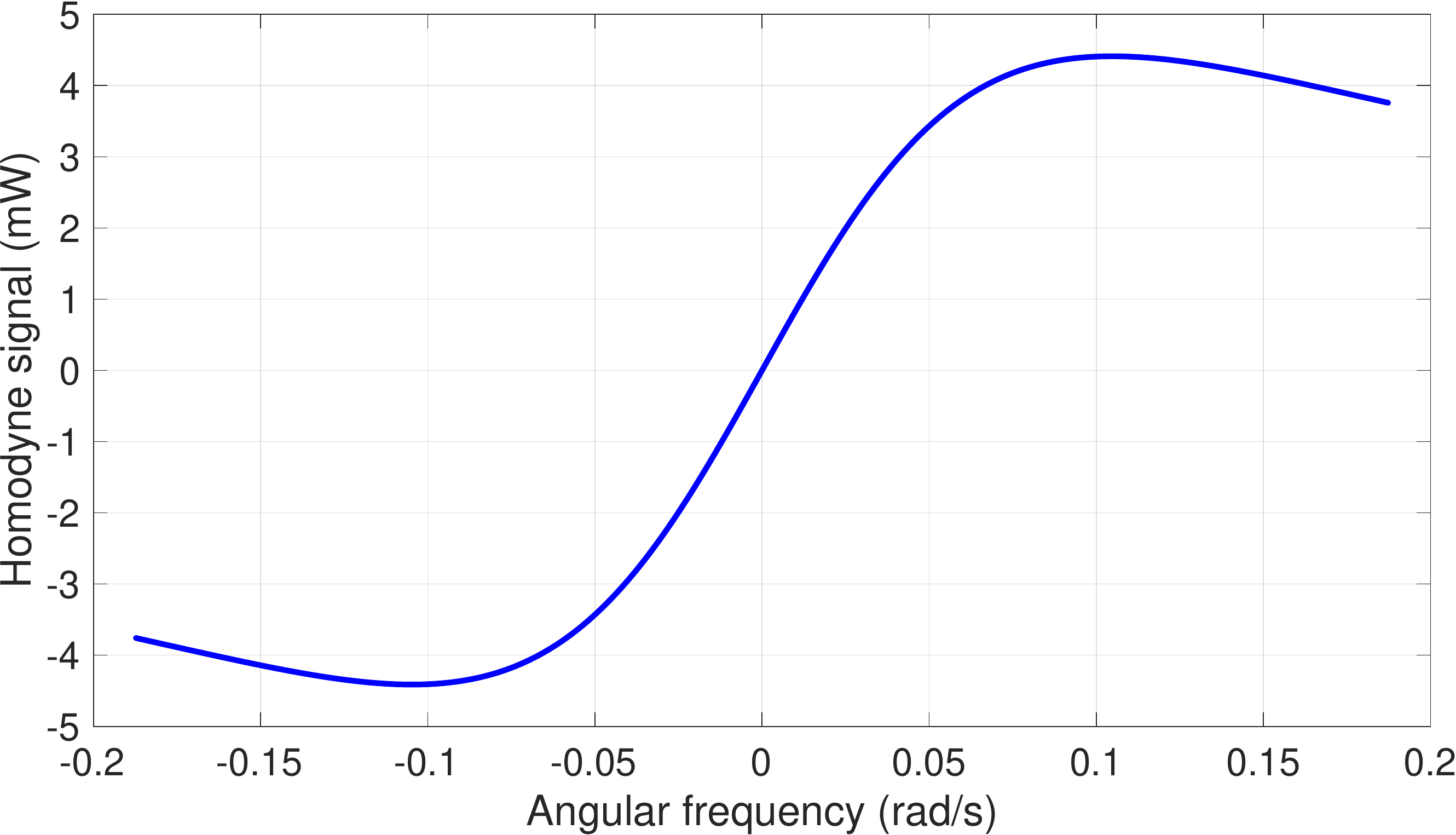}
        \caption{Response of the gyroscope to rotation of the frame. There is no lock-in behaviour at small angular frequencies. The range of the gyroscope is limited due to the open loop operation of the rotation channel. 
        %The range can be increased by normalising rotation readout by the intra-cavity power level.
        }
        \label{fig:range}
\end{figure}

Homodyne signals in the far detuned regime are simulated using Optickle~\cite{Optickle} software and are shown in Fig.~\ref{fig:homodyne} and~Fig.~\ref{fig:range}. The laser frequency is actively controlled with a bandwidth of 50\,kHz to keep $K_{\rm sym}$ around zero. We note that the derivative of the homodyne signal $K_{\rm sym}$ over the frequency detuning $\Delta f$ has different signs for the resonances of S- and P-polarisations. Therefore, the sign of the feedback control servo determines which mode resonates in the cavity. 

In our experiment, small ground vibrations determine the rotation of the optical table and the gyroscope. Therefore, the gyroscope stays in the linear regime. However, there is an angular velocity when eigen frequencies of the two counter-propagating beams in the optical cavity shift by a significant amount and the frequency servo can no longer sustain resonance conditions. The threshold angular velocity is given by the equation
\begin{equation}
	\Omega_{\rm max} =  \frac{\lambda c}{4 \pi B A} = 0.1
		  \left( \frac{20\,{\rm cm}}{a} \right)^2 \left( \frac{B}{6500} \right) \frac{\rm rad}{\rm s}.
\end{equation}

\begin{figure}[!t]
        \includegraphics[width=0.9\columnwidth]{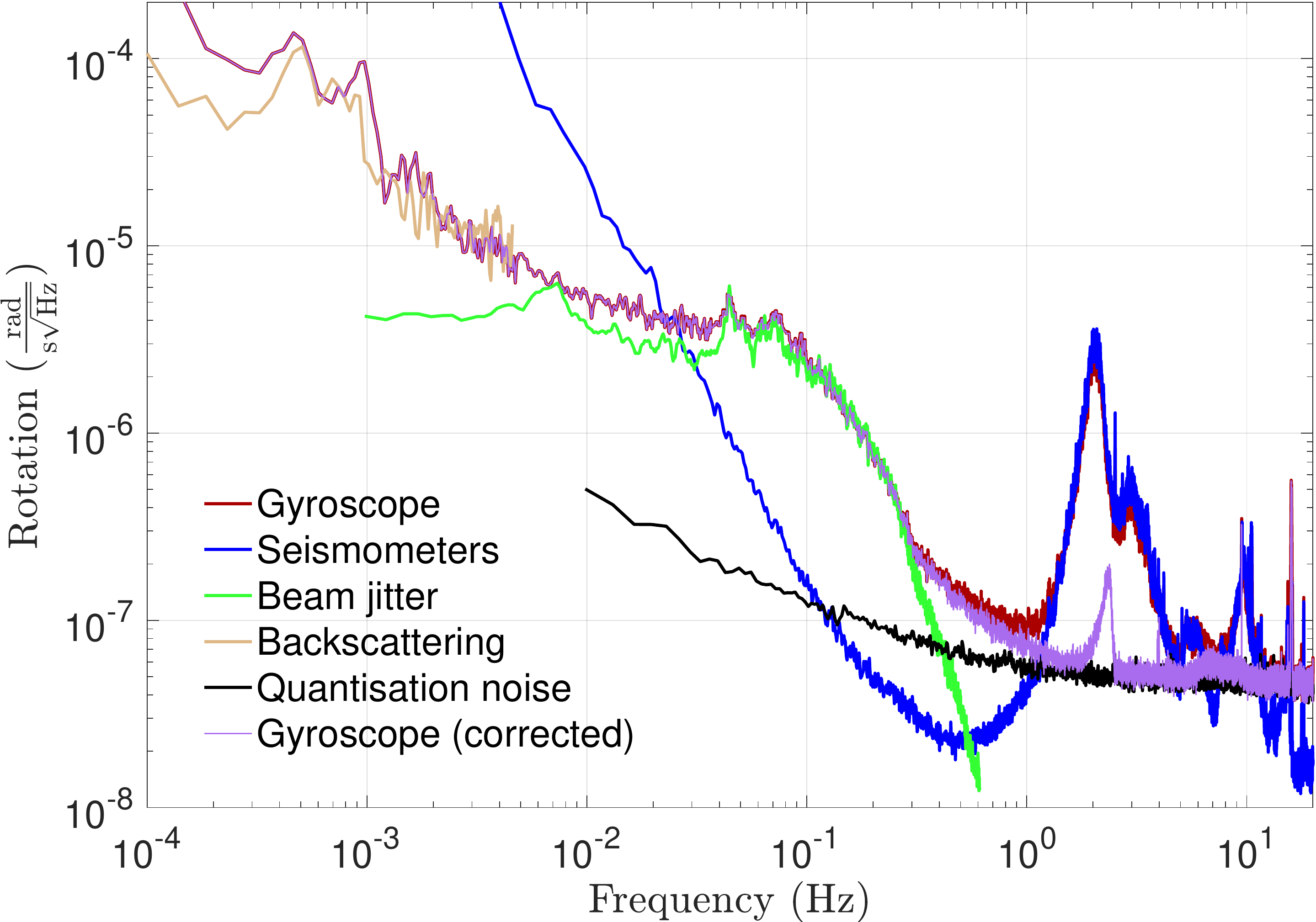}
        \caption{The sensitivity of the gyroscope and estimated noise sources.}
        \label{fig:NB}
\end{figure}

Fig. \ref{fig:NB} shows the sensitivity of the gyroscope and couplings of the noise sources in units of rad/s/$\sqrt{\rm Hz}$ which are common in the gravitational-wave community. The readout channel was calibrated from W to rad/s using Eq.~\ref{eq:cal_as}. We have also verified our calibration using a pair of Guralp 40T seismometers which were separated by 1.3 meters and measured rotation of the optical table. Above 1\,Hz, the gyroscope is limited by the motion of the optical table which is passively damped using pneumatic vibration isolators. These isolators are responsible for mechanical resonances around 2\,Hz and 9\,Hz which are clearly seen in the gyroscope and differential seismometer channels. In order to estimate the gyroscope noise hidden below the table motion, we corrected the gyroscope signals in the 1-10\,Hz band according to the equation
\begin{equation}
	S_{\rm corr} = S_{\rm gyro} - \frac{\left| C_{\rm gyro, sei} \right|^2}{S_{\rm sei}},
\end{equation}
where $C_{\rm gyro, sei}$ is the cross-spectral density between the gyroscope and seismometer rotation channels, $S_{\rm gyro}$ and $S_{\rm sei}$ are power spectral densities of the gyroscope and seismometer signals.

Photon shot noise and electronics noise cause sensing noises of the gyroscope readout channel. The power spectral density of the shot noise is dominated by the optical power in the P-polarisation: $P_{\rm as, p} = (T_p - R_p)^2 P_{\rm in, p}$ and $S_{\rm shot, W} = 2 h \nu P_{\rm as, p}$ in units of W$^2$/Hz, where $h$ is the Planck constant and $\nu = c / \lambda$ is the frequency of light. This spectrum is converted to units of rotation using Eq.~\ref{eq:cal_as} and is given by the equation
\begin{equation}
           \sqrt{S_{\rm shot}} = 7 \cdot 10^{-10} 
                \sqrt{\frac{4\,{\rm mW}}{P_{\rm in, s}}}  \left( \frac{10\,{\rm cm}}{a} \right)^2
                 \left( \frac{6500}{B} \right) \frac{\rm rad/s}{\sqrt{\rm Hz}}.
\end{equation}
The noise floor of our gyroscope is two orders of magnitude above the shot noise above 1\,Hz and is limited by the quantisation noise from the analog-to-digital converter which digitises the gyroscope rotation signal. Noise from the analog-to-digital converter can be improved using whitening filters in the future versions of the experiment.
 
Below 1\,Hz, air currents and backscattering in the optical cavity limit the sensitivity of the gyroscope. Air currents cause spatial jitter of the beam and fluctuations of the index of refraction. In the ideal case, our setup is insensitive to beam motion or jitter in the first order, but alignment drifts over time (due to low-frequency temperature fluctuations inside the isolation box) cause coupling to the rotation readout. We reduce beam jitter by enclosing our cavity beam with tubes and confining the optics in custom packages. We estimated coupling of the air currents by measuring the beam jitter in transmission of the input mirror using a quadruple photodetector (see Fig.~\ref{fig:diagram}). The beam jitter noise limits our sensitivity in the $30-500$\,mHz band.

Backscattering in the ring cavities has been studied as the  
cause of lock-in behaviour in active and passive~\cite{Zarinetchi:86} systems when the rotation
degree of freedom is controlled. In our system, backscattering from the counter-propagating cavity beams to each other
causes extra noise since cavity mirrors move relative to each other and only the total cavity length is controlled. The dominant source of the cavity mirror motion below 1\,Hz is thermal expansion of the gyroscope support blade which is made out of aluminum with a linear thermal expansion coefficient $\alpha = 1.1 \times 10^{-5}$\,/K. The isolation box provides passive filtering of the ambient temperature fluctuations with a time-scale of two hours. Residual temperature fluctuations $\Delta T$ couple to the gyroscope rotation channel according to the equation 
\begin{equation}\label{eq:sc_noise}
	\sqrt{S_{\rm sc}} = 10^3  \sqrt{\frac{R}{10^{-11}}} \sqrt{S_{\rm long}},
\end{equation}
where $R = {\rm BRDF}\times \lambda^2/ (\pi w^2)$ is reflectivity of the mirror to the counter-propagating mode, BRDF is bidirectional scattering distribution function of the cavity mirrors, $w$ is the beam size on the mirrors and $\sqrt{S_{\rm long}} = \alpha \Delta T a$ is the amplitude spectral density of the longitudinal motion of the mirrors. We measured the temperature fluctuations in the isolation box using a set of thermistors and estimated the noise from backscattering according to Eq.~\ref{eq:sc_noise}. 
%The corresponding trace is shown in Fig.~\ref{fig:NB}.

\begin{figure}[!t]
        \includegraphics[width=0.9\columnwidth]{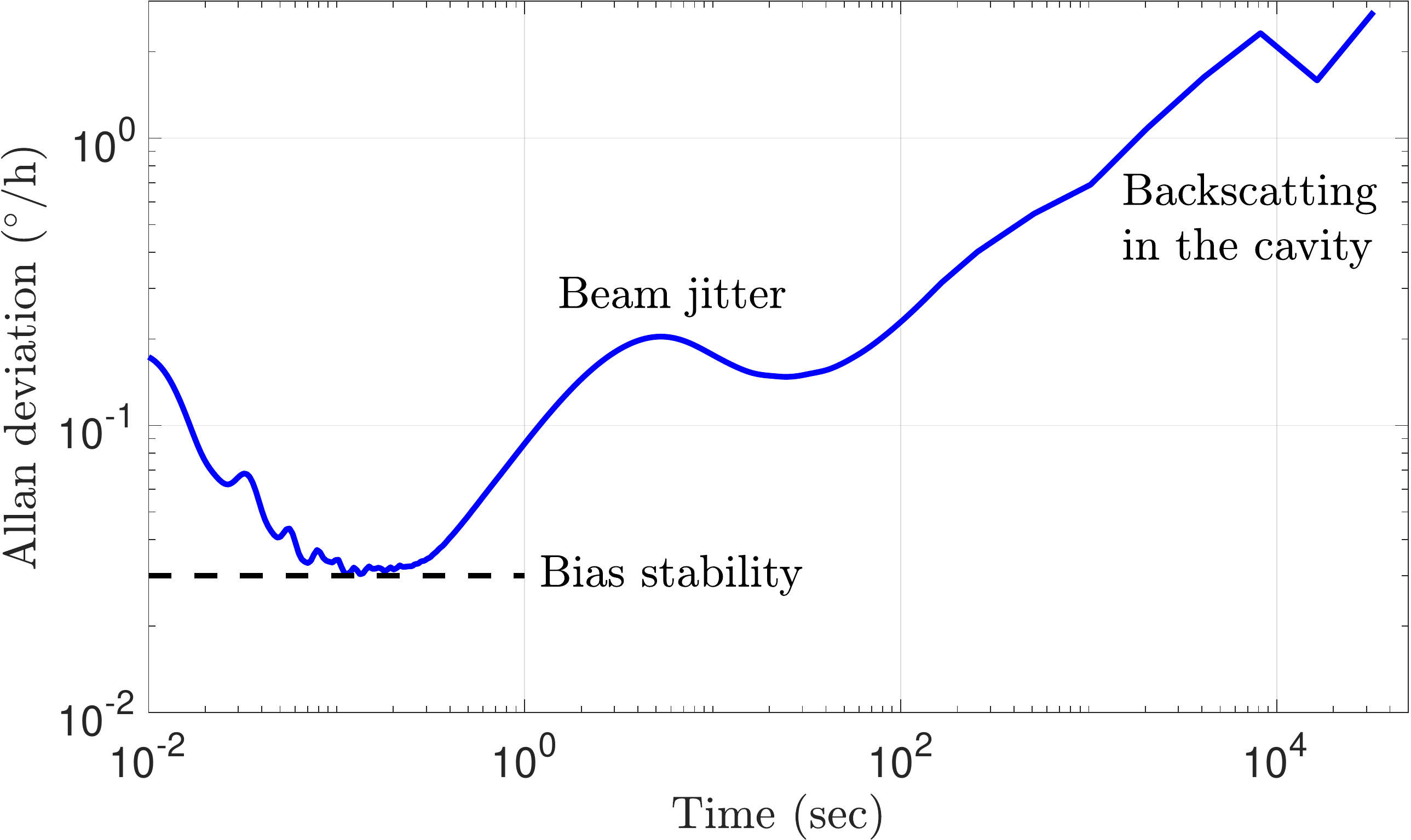}
        \caption{The Allan deviation of the gyroscope.}
        \label{fig:dev}
\end{figure}

We have also measured coupling of the lasers frequency and amplitude noises. These noises couple to the gyroscope readout through imbalances in the system, such as backscattering in the optical cavity and misalignments of the input optics. We simulated coupling from the lasers noises in the Optickle software and compared the simulations with measured transfer functions. We conclude that laser noises are insignificant noise sources in our experiment.

Fig.~\ref{fig:dev} shows the Allan deviation of the rotation channel, a common metric of navigation gyroscope performance, in units of $^\circ$\!/h. For small times $\tau<0.1$\,sec the deviation improves as $\sqrt{\tau}$ since electronics noise is white above 1\,Hz. We reach bias stability of 0.03\,$^\circ$/h at $\tau = 0.15$\,sec. At longer times, the Allan deviation increases due to the beam jitter noise and backscattering in the optical cavity. On the time scales of a few hours, the deviation is dominated by temperature fluctuations of the gyroscope.

In conclusion, we have demonstrated a novel design for a passive gyroscope, utilising a
folded, free-space, resonant cavity. We have shown control of the gyroscope using a double homodyne readout technique. Although this paper presents a design with a free-space cavity, this
method of readout can be applied to any cavity that exhibits a phase shift between its eigen polarisations, opening this
technique's applicability to a wider scope of resonators. In this experiment, we reached the sensitivity of $5 \times 10^{-8}$\,\sens above 1\,Hz and are limited by the table motion, electronics noise of the analog-to-digital converter at these frequencies. Below 1\,Hz we are limited by the beam jitter and thermal motion of the cavity mirrors which couple to the rotation channel due to backscattering in the optical cavity.

The next step will be to reduce the motion of the cavity mirrors using passive and active stabilisation. In particular, we plan to build a cavity on a monolithic fused silica plate similar to the LISA interferometers~\cite{Gerberding_2017}. Since the thermal expansion coefficient of fused silica is a factor of 30 smaller compared to aluminum, we expect reduction of the cavity mirror motion by the same factor. Furthermore, we plan to actively stabilise the gyroscope temperature with a system similar to the stabilisation in atomic clocks and compact gyroscopes which reach the thermal stability of $10^{-5}$\,K over several hours~\cite{Liu:16}. Active temperature control has a potential to reduce the cavity motion by two orders of magnitude compared to the current design and significantly reduce the noise level around 1\,mHz.

\textbf{Funding.} The Kavli Foundation.
M.E. acknowledges 
the support of the National Science Foundation 
and the LIGO Laboratory. 
LIGO was constructed by the California Institute
 of Technology and Massachusetts Institute of Technology with funding
 from the National Science Foundation and operates under cooperative
 agreement PHY-0757058.

%The authors thank H. Haase, C. Wiede, and J. Gabler for technical support.
%
%\section{References}
%
%Note that \emph{Optics Letters} uses an abbreviated reference style. Citations to journal articles should omit the article title and final page number; this abbreviated reference style is produced automatically when the \emph{Optics Letters} journal option is selected in the template, if you are using a .bib file for your references. 
%
%However, full references (to aid the editor and reviewers) must be included as well on a fifth informational page that will not count against page length; again this will be produced automatically if you are using a .bib file. 
%
%\bigskip
%\noindent Add citations manually or use BibTeX. See \cite{Zhang:14,OSA,FORSTER2007}.
%
% Bibliography
\bibliography{paper}

% Full bibliography added automatically for Optics Letters submissions
% Note that this extra page will not count against page length
\ifthenelse{\equal{\journalref}{ol}}{%
\clearpage
\bibliographyfullrefs{paper}
}{}
 
% Please include bios and photos of all authors for aop articles 

%\ifthenelse{\equal{\journalref}{aop}}{%
%\section*{Author Biographies}
%\begingroup
%\setlength\intextsep{0pt}
%\begin{minipage}[t][6.3cm][t]{1.0\textwidth} % Adjust height [6.3cm] as required for separation of bio photos.
%  \begin{wrapfigure}{L}{0.25\textwidth}
%    \includegraphics[width=0.25\textwidth]{john_smith.eps}
%  \end{wrapfigure}
%  \noindent
%  {\bfseries John Smith} received his BSc (Mathematics) in 2000 from The University of Maryland. His research interests include lasers and optics.
%\end{minipage}
%\begin{minipage}{1.0\textwidth}
%  \begin{wrapfigure}{L}{0.25\textwidth}
%    \includegraphics[width=0.25\textwidth]{alice_smith.eps}
%  \end{wrapfigure}
%  \noindent
%  {\bfseries Alice Smith} also received her BSc (Mathematics) in 2000 from The University of Maryland. Her research interests also include lasers and optics.
%\end{minipage}
%\endgroup
%}{}

\end{document}